%% file: Camera_ready_Arxiv.tex
\documentclass[conference]{IEEEtran}
\usepackage{graphicx,times,cite,amsmath, amssymb, epsfig, array, mathtools, epstopdf,makecell, algorithm, algorithmic,dsfont,color,xfrac}
\newlength{\figwidth}
\begin{document}
\newcommand{\new}[1]{\textcolor{blue}{#1}}	
\setlength{\pdfpagewidth}{8.5in}
\setlength{\pdfpageheight}{11in}

\newtheorem{theorem}{Theorem}
\newtheorem{corollary}{Corollary}
\newtheorem{lemma}{Lemma}
\newtheorem{proposition}{Proposition}
\newtheorem{condition}{Condition Set}
\newtheorem{prob}{Problem}

\title{Average Transmission Success Probability Bound for SWIPT Relay Networks}

\author{
	\IEEEauthorblockN{ Bhathiya Pilanawithana, Saman Atapattu and Jamie Evans}
	\IEEEauthorblockA{Department of Electrical and Electronic Engineering, University of Melbourne, Australia\\ e-mail: mpilanawitha@student.unimelb.edu.au, \{saman.atapattu, jse\}@unimelb.edu.au}}
\maketitle

\begin{abstract}
Wireless energy transferring technology offers a constant and instantaneous power for low-power applications such as Internet of Things (IoT) to become an affordable reality. This paper considers simultaneous wireless information and power transfer (SWIPT) over a dual-hop decode-and-forward (DF) relay network with the power-splitting (PS) energy harvesting protocol at the relay. The relay is equipped with a finite capacity battery. The system performance, which is characterized by the average success probability of source to destination transmission, is a function of the  resource allocation policy that selects the PS ratio and the transmit energy of the relay. We develop a mathematical framework to find an upper bound for the maximum the average success probability. The upper bound is formulated by a discrete state space Markov decision problem (MDP) and make use of a policy iteration algorithm to calculate it.
\end{abstract}

\begin{IEEEkeywords}
Power-splitting protocol, relay network, resource allocation, wireless energy transfer.
\end{IEEEkeywords}
\section{Introduction}
Multi-user networks with relays, sensors and Internet of Things (IoT) in the 5G and beyond networks will generate enormous amount of data and consume large amount of energy for a wide range of services in different domain, e.g.,  \cite{Andrews2014jsac,Atapattu2019twc} and references therein. 
One of the key challenges in such wireless networks is energizing the remote devices for successful communication. Although natural energy resources such as wind and solar can be used, they are often hindered by inconsistent availability, implementation overhead or the requirement of large infrastructure. Thus, energy harvesting (EH) using radio frequency (RF) signals, is motivated as existing communication circuitry can be used with low cost  modifications \cite{Zhou}. Since such low power communication interfaces make the seamless connectivity more challenging, relaying or cooperative communication has been promoted as a viable solution, especially for the Internet of Things (IoT)  \cite{Wang}. Thus, RF energy harvesting in relay networks has gained much attention recently. 

\subsection{Related Work}
Since energy at the EH node is not automatically replenished as in a traditional node with fixed power supply, the performance of an EH network depends on the EH protocol and the usage scheme of the  harvested energy. For simultaneous information and power transfer (SWIPT), two basic EH protocols, i) time-switching (TS) and  ii) power-splitting (PS), are introduced for amplify-and-forward (AF) and decode-and-forward (DF) relay networks in \cite{Nasir1,Saman1,Saman2}. An optimal hybrid EH protocol, which is a combination of PS and TS protocols is introduced in \cite{Saman3,bhathiya1} and it outperforms both TS and PS protocols. An improved receiver architecture for PS protocol is introduced in \cite{FTChien} and \cite{WChoi}, which makes use of the level of the harvested energy as side information to assist the decoding of the source transmitted message. The common assumption of most of these work is that the total harvested energy is used for data transmission and thus a battery for long term energy storage is not required at the EH node. However, a long term energy storage enables a PS energy harvesting node to manage two basic resources i) PS ratio and ii) transmit energy. Thus, an efficient resource allocation scheme, which store excess amount of harvested energy for future use, can achieve a better performance compared to a network without a battery in the EH node. Due to the battery energy dependency on the resource allocation decisions made earlier, the analysis of the system performance needs more attention.

For EH relaying with a battery, several resource allocation methods are discussed in literature. An AF relaying network with TS energy harvesting is considered in \cite{Krikidis}, where data relaying is realized when sufficient energy is collected through EH. An AF relaying network with PS energy harvesting is considered in \cite{Blum}, where the remaining energy after data transmission is stored in the battery. The optimal resource allocation that maximizes the energy efficiency in a WSN with DF relaying is considered in \cite{LZheng}. A sum-throughput maximization problem is formulated for DF relay \cite{WTu}, where the relay node opportunistically switch between modes of total EH and PS based information processing. Resource allocation schemes for EH nodes which harvest energy from renewable sources such as wind or solar are investigated in \cite{FYuan, Amirnavaei}. All these work assume full CSI at the decision node. The outage performance is analyzed in \cite{HJiang} for a sub-optimal resource allocation scheme based on incremental DF relay protocol.

\subsection{Problem Statement and Contribution}
In contrast to previous work \cite{Nasir1,Saman1,Saman2,Saman3,Blum,Krikidis,LZheng,WTu}, this paper thus considers a dual hop DF relaying network with the PS energy harvesting protocol assuming that no CSI of forward channels is available at any node. The system performance is evaluated by the average success probability of the source to destination communication. To efficiently use the harvested energy, the relay is equipped with a battery, which consists of a finite capacity. In contrast to \cite{HJiang}, we focus our attention to find the maximum average success probability over the set of resource allocation policies. The evaluation of  maximum is important to assess the feasibility of the network for a practical set of system parameters. Due to the intractability of the problem, we develop a mathematical framework to find an upper bound for the maximum average success probability by formulating a discrete state Markov decision problem (MDP).

\section{System Model}\label{system_model}

\begin{table}[]
	\centering
	\caption{Notations}
	\label{notations}
	\renewcommand{\arraystretch}{1}
	
	\begin{tabular}{| >{\centering\arraybackslash}m{2.3cm}| >{\centering\arraybackslash}m{5.5cm}|}
		\hline
		Notation & Remark   \\ 
		\hline\hline
		$P_s$ & Source transmit power   \\
		\hline 
		$\sigma^2$ & Noise power   \\
		\hline 
		$T$ & Block duration   \\
		\hline 
		\small{$m$} & \small{Block index}   \\
		\hline 
		\small{$h_m$} & \small{S-R channel power gain in the $m$th~block}   \\
		\hline
		\small{$g_m$} & \small{R-D channel power gain in the $m$th~block}  \\
		\hline
		\small{$E_m$} & \small{Battery energy at the beginning of the $m$th~block}   \\
		\hline
		\small{$\lambda_m$} & \small{PS ratio used in the $m$th~block}   \\
		\hline
		\small{$u_m$} & \small{Relay transmit energy used in the $m$th~block}   \\
		\hline
		\small{$S_m$} & \small{State of the relay in the $m$th~block - $\left(E_m,h_m\right)$ pair } \\
		\hline
		\small{$A_m$} & \small{Relay action in the $m$th~block - $\left(\lambda_m,u_m\right)$ pair}  \\
		\hline
		\small{$\mathcal{S}$} & \small{State space - set of all possible $S_m$}  \\
		\hline
		\small{$\mathcal{A}_s$} & \small{Action space - set of all possible $A_m$} \\
		\hline
		\small{$d_m\left(\cdot\right)$} & \small{Decision rule in the $m$th~block, which gives an action for each state - $A_m=d_m\left(S_m\right)$}\\
		\hline
		\small{$\pi$} & \small{Resource allocation policy - the sequence of decision rules $d_1,d_2,\cdots$}\\
		\hline
		\small{$\widetilde{\text{P}}_\pi\left(s\right)$} & \small{Average success probability of policy $\pi$ for the initial state $S_1=s$} \\
		\hline
		\small{$\text{P}_\pi$} & \small{Average success probability of policy $\pi$} \\
		\hline
	\end{tabular}
\end{table}

In this section, we discuss main assumptions and the operation of the network.
\subsection{Network Model}
We consider a wireless relay network in which a source node (S) communicates with a destination node (D) via a single relay node (R). The relay operates in the DF mode. We assume that the direct link between S and D is not available due to a blockage. The communication takes place in half-duplex mode. Each node has a single antenna. 

The network operates block by block, where each block has a duration $T$ and is indexed by $m \in \{1,2,\cdots\}$. The fading coefficients of S to R channel (S-R) and R to D channel (R-D) in the $m$th~block are denoted by $\tilde{h}_m$ and $\tilde{g}_m$, respectively, which are independent. Since an unbounded flat-fading channel may be modeled by a finite number of channel states with an arbitrary low error \cite{Parastoo, Blum}, both channel coefficients are drawn from finite sets. We assume that there is no feedback from D to R or from R to S. Thus, no CSI is available on the forward channel, i.e., S does not have any channel knowledge, R has knowledge on $\tilde{h}_m$, and D has knowledge on $\tilde{g}_m$. The source transmits with constant power $P_s$ and information rate $\tau$. The relay harvests energy from source transmitted information signal and uses that energy for information transmission to the destination. The PS protocol is used in R. The source transmits the message during the first half of the block. The relay uses $\sqrt{\lambda_m}$ portion of the received signal for the EH, and the remaining $\sqrt{1-\lambda_m}$ portion of the received signal is utilized for the information decoding. During the second half of the block, the relay transmits the decoded message to the destination using $u_m$ amount of energy.

\subsection{Analytical Model}

\subsubsection{S-R and R-D Transmission}
The discrete time received signal at the information decoder of $R$ in $k$th symbol index of $m$th block is
\begin{equation*}
\hat{y}_{r,m}^{(k)} = \sqrt{1-\lambda_m}\bigg(\sqrt{P_s}  \tilde{h}_{m} s^{(k)}_m + n_{r,a}^{(k)}\bigg)+n_{r,c}^{(k)} \, ,
\end{equation*}
where $s^{(k)}_m$ is the $k$th symbol transmitted by S, $n_{r,a}^{(k)}$ and $n_{r,c}^{(k)}$ are AWGN at the antenna and the information decoder of $R$, respectively with variance $\sigma^2$. Therefore, the signal-to-noise-ratio (SNR) of $S-R$ channel in the $m$th block is
\begin{equation}\label{gamSR}
\gamma_1\left(h_m,\lambda_m\right)= \frac{\left(1-\lambda_m\right) h_m P_s}{\left(2-\lambda_m\right) \sigma^2} \, , 
\end{equation}
where $h_m= | \tilde{h}_m|^2$ and $h_m \in \mathcal{H}$ for all $m$. Since fading coefficients are drawn from a finite set, $\mathcal{H}$ is also finite. Thus, we have $\mathcal{H}=\left\{h^{(i)}| \ i=1,2,\cdots,N_c \right\}$, where $N_c$ is the total number of elements in $\mathcal{H}$. To omit the use of the index $i$ when not necessary, we may denote a general element of $\mathcal{H}$ by $h$. The probability mass functions for $\mathcal{H}$ is $f_\mathcal{H}\left(h\right)$.

If the Relay uses $u_m$ energy to transmit information, the discrete time received signal at $D$ in the $k$th symbol index of the $m$th block is
\begin{equation*}
{y}_{d,m}^{(k)} = \sqrt{\frac{2  u_m}{T}} \tilde{g}_{m} \hat{s}^{(k)}_m + n_{d,a}^{(k)} + n_{d,c}^{(k)}\, ,
\end{equation*}
where $\hat{s}^{(k)}_m$ is the $k$th symbol transmitted by R. Therefore the SNR at D in the $m$th~block is
\begin{equation}\label{gamRD}
\gamma_2\left(g_m,u_m\right) = \frac{u_m  g_m}{T \sigma^2} \, , 
\end{equation}
where $g_m= | \tilde{g}_m|^2$ and $g_m \in \mathcal{G}$ for all $m$. Since fading coefficients are drawn from a finite set, $\mathcal{G}$ is also finite. We denote the largest element of $\mathcal{G}$ by $g_{max}$.

\subsection{Relay Operations and Battery Behavior}\label{relay_operations}
The total harvested energy during the $m$th~block by neglecting the noise energy, is $\eta P_s h_m \lambda_m \frac{T}{2}$ where $\eta\in (0,1)$ is the conversion efficiency \cite{Zhou}. This energy is directly transfered  to the battery. Thus, the battery energy at $t=\left(m+\frac{1}{2}\right)T$ is
\begin{equation}\label{diff_eqn_sol}
E_{m+\frac{1}{2}} = \text{min}\Bigg[ \frac{\eta P_s h_m \lambda_m T}{2} + E_m , B\Bigg]  \, ,
\end{equation}
where $B<\infty$ is the battery capacity and $E_m$ is the residual battery energy at the beginning of the $m$th~block. 

For information transmission from R to D, the relay uses $u_m$ amount energy. The residual battery energy for the next block, is
\begin{equation}\label{end_block_energy}
E_{m+1} = \left[E_{m+\frac{1}{2}} - u_m\right] \, .
\end{equation}

If Shannon channel capacity is larger than the information rate $\tau$, the receiving node may decode the received signal with arbitrary small error probability. This is defined as a successful decoding. Thus, to achieve a successful decoding with a minimum received SNR $\gamma_\tau$, we have $\tau = \frac{1}{2}\mathrm{log}_2\left(1+\gamma_\tau\right)$\,bits/s/Hz, in which the factor $\frac{1}{2}$ is due to each S-R and R-D links are used only half of the total time. This satisfies $\gamma_\tau = 4^\tau -1$. Thus, for a successful decoding at the relay and the destination, we have $\gamma_1 \left(h_m,\lambda_m\right) \geqslant \gamma_\tau$ and $\gamma_2 \left(g_m,u_m\right) \geqslant \gamma_\tau$, respectively. The PS ratio $\lambda_m$ and relay transmit energy $u_m$ used, impact the SNRs $\gamma_1\left(h_m,\lambda_m\right)$ and $\gamma_2\left(g_m,u_m\right)$. Subsequently, they effect the probability of successful transmission from the source to the destination.  In the next section, we discuss the calculation of the average success probability.

\section{The Average Success Probability}\label{succ_cal_section}

We first define the \emph{state} $S_m$ in the $m$th~block to be the pair $S_m=\left(E_m, h_m\right)$. The state $S_m$ for each $m$, takes an element from the the \emph{state space} defined as\break $\mathcal{S}=\left\{s=\left(E,h\right)|\ h\in \mathcal{H}, E\in [0,B] \right\}$, where a general element of $\mathcal{S}$ is denoted by $s=\left(E,h\right)$. The \emph{action}, $A_m$, taken by the relay in the $m$th~block is defined as the pair $A_m=\left(\lambda_m, u_m\right)$. For the brevity, we then define  two functions related to \eqref{diff_eqn_sol} and \eqref{end_block_energy} as 
\begin{equation}\label{diff_eqn_sol2}\begin{split}
\mathcal{E}_{\frac{T}{2}}\left(\lambda_m,E_m,h_m\right) &= \text{min}\Bigg[ \frac{\eta P_s h_m \lambda_m T}{2} + E_m , B\Bigg] \, , \\
\mathcal{E}_{T}\left(\lambda_m,u_m,E_m,h_m\right) &= \left[\mathcal{E}_{\frac{T}{2}}\left(\lambda_m,E_m,h_m\right) - u_m\right]\, ,
\end{split}
\end{equation}
which are used to represent $E_{m+\frac{1}{2}}= \mathcal{E}_{\frac{T}{2}}\left(\lambda_m,E_m,h_m\right)$ and $E_{m+1}=\mathcal{E}_{T}\left(\lambda_m,u_m,E_m,h_m\right)$, respectively. The PS ratio $\lambda_m$ may take any value in $[0,1]$. The transmit energy $u_m$ and the residual battery energy for the next block $E_{m+1}$ are non-negative. By considering these constraints, the action $A_m$ at each $m$ takes an element from the \emph{action space}, $\mathcal{A}_s$, which is defined as the set of all actions for state $s$ and it can be given as
\begin{equation}\label{As}
\mathcal{A}_s=\left\{a=\left(\lambda, u\right) |\ \lambda\in [0,1]  , \  0\leqslant u \ , \ 0\leqslant \mathcal{E}_{T}\left(\lambda,u,s\right)\right\} \, , 
\end{equation}
where a  general element of $\mathcal{A}_s$ is denoted by $a=\left(\lambda,u\right)$. 

The knowledge of $S_m=\left(E_m, h_m\right)$ is available in the relay at the beginning of each $m$th~block. We thus consider each action $A_m$ as a function of the current state denoted by $d:\mathcal{S}\to \mathcal{A}_s$, i.e. $A_m = d\left(S_m\right)$, where this function is termed as the \emph{decision rule}. Since each action is an element of $\mathcal{A}_s$, the \emph{decision rule space}, $\mathcal{D}$, which is the set of all possible decision rules can be given as
\begin{equation}\label{D_space}
\mathcal{D}= \left\{d \ | \ d\left(s\right) \in \mathcal{A}_s \forall s \in \mathcal{S} \right\} \ .
\end{equation}
The relay can be configured to have a sequence of decision rules $\pi=\left\{d_1, d_2,\cdots \right\}$, which is termed as \emph{policy}. For each $S_m$, the action $A_m$ is chosen according to $d_m$. The \emph{policy space} is thus given by $\Pi = \mathcal{D}\times\mathcal{D}\times\mathcal{D}\times\cdots$. A stationary policy employs the same decision rule $d$ at all blocks, i.e., $d^\infty$. Without loss of generality, we may denote a stationary policy by $d$.

For a given state $S_m=\left(E_m, h_m\right)$ and action $A_m=\left(\lambda_m,u_m\right)$, the success probability of S-R link can be given as 
\begin{align}\label{suc_prob_SR_for_given_Et_gSRt}
\Pr\left(\text{S-R} \ \text{success} \, \big|\, S_m,A_m\right) &\overset{(a)}{=}\mathds{1}_{\left[\gamma_1\left(h_m,\lambda_m\right) \geqslant \gamma_\tau\right]} \nonumber \\
&\overset{(b)}{=} \mathds{1}_{\left[\lambda_m \leqslant \frac{h_m P_s - 2 \sigma^2 \gamma_\tau}{h_m P_s -  \sigma^2 \gamma_\tau}\right]} \, ,
\end{align}
where $\mathds{1}_{\left[\gamma_1\left(h_m,\lambda_m\right) \geqslant \gamma_\tau\right]}=1$ when $\gamma_1\left(h_m,\lambda_m\right) \geqslant \gamma_\tau$, and $0$ otherwise. The equation $(a)$ follows as the requirements for the successful decoding at the relay, and $(b)$ comes from (\ref{gamSR}). For a given state $S_m=\left(E_m, h_m\right)$, and action $A_m=\left(\lambda_m,u_m\right)$, the success probability in R-D link can be given with the aid of \eqref{gamRD} as 
\begin{align}\label{suc_prob_RD_for_given_Et_gSRt}
\Pr\left(\text{R-D} \ \text{success} \, \big|\, S_m,A_m \right) &= \Pr\left(g_m\geqslant\frac{T  \sigma^2 \gamma_\tau}{u_m}\right) \, .
\end{align}
For state $S_m$ and action $A_m$, we define the \emph{reward}, $p\left(S_m, A_m\right)$,  as the end-to-end success probability, which is evaluated as  
\begin{equation}\label{suc_prob_endtoend_blockn}
\begin{split}
p\left(S_m, A_m\right) =\Pr\left(g_m\geqslant\frac{T  \sigma^2 \gamma_\tau}{u_m}\right) \mathds{1}_{\left[\lambda_m \leqslant \frac{h_m P_s - 2 \sigma^2 \gamma_\tau}{h_m P_s -  \sigma^2 \gamma_\tau}\right]} \, .
\end{split}
\end{equation}

For the policy $\pi=\{d_1,d_2\cdots\}$ and the initial state $S_1=s$, the time average success probability over $M$ blocks is given as 
\begin{equation}\label{lim_avg_suc_prob}
\bar{p}_{\pi,M}\left(s\right) = \frac{1}{M} \mathbb{E}\left[\sum_{m=1}^{M} p\left(S_m, d_m\left(S_m\right)\right) \, \bigg| \,S_1=s\right] \, ,
\end{equation}
where $\mathbb{E}\left[\cdot\right]$ denotes the 
expectation operator. The long term average success probability for initial state $S_1=s$, is thus given by $\widetilde{\text{P}}_\pi\left(s\right)=\lim_{M \to \infty} \bar{p}_{\pi,M}\left(s\right)$. We consider all policies for which the limit exists. Without loss of generality, we assume that the initial battery energy $E_1=0$. The channel fading is independant from the battery energy in the relay. Therefore, the long term average success probability is given by
\begin{equation}\label{lim_inf_avg_suc_prob_vector}
\text{P}_\pi = \mathbb{E}\left[\widetilde{\text{P}}_\pi\big(\left(0,h\right)\big)\right] \, .
\end{equation}

It is important to find the maximum $\text{P}_\pi$ in order to assess the feasibility of the system. Since the state space $\mathcal{S}$ and the action space $\mathcal{A}_s$ is uncountably infinite, maximization of  $\text{P}_\pi$ with respect to policy $\pi$, is intractable. Therefore, the main objective of this paper is to find an upper bound for the maximum $\text{P}_\pi$, denoted by $P_u $, by making use of a suitable discretization of $\mathcal{S}$ and $\mathcal{A}_s$. For comparison purposes we also provide a heuristic resource allocation policy. These will be discussed in the next section

\section{A Heuristic Policy and the Upper bound}\label{Upper_Lower_bound_calculation}

We notice that in some states $s\in \mathcal{S}$ any action $a\in \mathcal{A}_s$ taken results in $p\left(s, a\right)=0$. Therefore, when deriving the heuristic policy and the upper bound $P_u$, these states can be treated differently to other states. To this end, we categories each state $s=\left(E,h\right)$ in to two subsets depending on the resulting reward $p\left(s, a\right)$ for action $a=\left(\lambda, u\right)$;
\begin{itemize}
	\item \emph{Subset-1} : $\mathcal{C}_1 = \Big\{\left(h, E\right)\in \mathcal{S} \, | \, \mathcal{E}_{\frac{T}{2}}\left(\frac{h P_s - 2 \sigma^2 \gamma_\tau}{h P_s -  \sigma^2 \gamma_\tau},h,E\right) <\frac{T \sigma^2 \gamma_\tau}{g_{max}} \ \text{or} \  h < \frac{2 \sigma^2 \gamma_\tau}{P_s} \Big\}$
	\vspace{0.4cm}
	
	As given in \eqref{suc_prob_SR_for_given_Et_gSRt}, when $\lambda > \frac{h P_s - 2 \sigma^2 \gamma_\tau}{h P_s -  \sigma^2 \gamma_\tau}$, the relay cannot decode the source message.  The maximum $\lambda$, which helps successful decoding is $\lambda = \frac{h P_s - 2 \sigma^2 \gamma_\tau}{h P_s -  \sigma^2 \gamma_\tau}$. The condition $h < \frac{2 \sigma^2 \gamma_\tau}{P_s}$ describes the situation where no $\lambda \in [0,1]$ satisfies $\lambda \leqslant \frac{h P_s - 2 \sigma^2 \gamma_\tau}{h P_s -  \sigma^2 \gamma_\tau}$, which causes $p\left(s, a\right)=0$ for all $a \in \mathcal{A}_s$.
	
	\quad On the other hand, it can be seen from (\ref{As}) that selection of $\lambda$ restricts the selection of $u$. A lager value for $\lambda$ allows the relay to harvest more energy, which results in more energy in the battery. This enable the relay to use a larger $u$. Therefore, with the aid of \eqref{end_block_energy}, the maximum value $u$ can take, while allowing the relay to decode the source message is $u= \mathcal{E}_{\frac{T}{2}}\left(\frac{h P_s - 2 \sigma^2 \gamma_\tau}{h P_s -  \sigma^2 \gamma_\tau},h,E\right)$. When the relay uses this energy to transmit to the destination, the largest SNR at the destination is achieved when $g=g_{max}$ in (\ref{gamRD}). The condition $\mathcal{E}_{\frac{T}{2}}\left(\frac{h P_s - 2 \sigma^2 \gamma_\tau}{h P_s -  \sigma^2 \gamma_\tau},h,E\right)<\frac{T \sigma^2 \gamma_\tau}{g_{max}}$ describes the situation when the largest achievable SNR falls below $\gamma_\tau$. This causes $p\left(s, a\right)=0$ for all $a \in \mathcal{A}_s$.
	
	Therefore, $p\left(s, a\right)=0$ for all $a \in \mathcal{A}_s$ whenever $s\in \mathcal{C}_1$.
	\vspace{0.4cm}
	
	\item \emph{Subset-2} : $\mathcal{C}_2=\mathcal{S}\backslash \mathcal{C}_1$
	\vspace{0.2cm}
	
	When the state $s$ does not belong to $\mathcal{C}_1$, we have $\mathcal{E}_{\frac{T}{2}}\left(\frac{h P_s - 2 \sigma^2 \gamma_\tau}{h P_s -  \sigma^2 \gamma_\tau},h,E\right) >0$, which makes \break $\lambda=\frac{h P_s - 2 \sigma^2 \gamma_\tau}{h P_s -  \sigma^2 \gamma_\tau}$ and $u> 0$ feasible. Therefore, whenever $s\in \mathcal{C}_2$, there exists an action $a \in \mathcal{A}_s$, which gives $p\left(s, a\right)>0$.
\end{itemize}

\subsection{Heuristic Policy}
If the conditional distribution of the state $S_{m+1}$ given $S_m=s=\left(h,E\right)$ is known, the evaluation of expectation operation in \eqref{lim_avg_suc_prob} is straight forward. A simple way this can be achieved is by driving the energy level of the battery to zero by using the total amount of the battery energy for $u_m$. Thus, for any $S_m$, the residual battery energy $E_{m+1}=0$ and the $h_{m+1}$ is independent from $S_m$. With the aid of \eqref{As}, a heuristic decision rule, which always drives the battery energy to zero can be given as
\begin{equation}\label{decision_rule_IRM}
d_{l}\left(s\right) = \begin{cases}
\lambda=1, \qquad\qquad\qquad\qquad\quad   \text{if} \ s \in \mathcal{C}_1\\ \quad u=\mathcal{E}_{\frac{T}{2}}\left(\lambda,h,E\right) \\ \lambda=\frac{h P_s - 2 \sigma^2 \gamma_\tau}{h P_s -  \sigma^2 \gamma_\tau}, \qquad\qquad\quad \text{otherwise \, .}\\ \quad u= \left[\mathcal{E}_{\frac{T}{2}}\left(\lambda,h,E\right) \right] 
\end{cases}
\end{equation}
The stationary policy generated by the above decision rule is $\pi_l=d_l^\infty$. If $\pi_l$ is used, the states $S_m$ for all $m>1$ is known to be an element from the set $\left\{\left(0,h\right)|\ h\in \mathcal{H} \right\}$. Therefore, the average success probability for initial state $S_1=s$ can be written as
{\small
\begin{multline*}
\widetilde{\text{P}}_{\pi_l}(s)=\lim_{M \to \infty} \frac{1}{M} \Bigg[ p\left(s,d_l\left(s\right)\right) 
+ \sum_{m=2}^{M} \mathbb{E}\left[p\left(\left(0,h\right), d_l\left(0,h\right)\right)\right]\Bigg] \ .
\end{multline*}
}
By taking the limit in the above equation and noting that $\widetilde{\text{P}}_{\pi_l}(s)$ is constant with respect to $s$, with the aid of \eqref{lim_inf_avg_suc_prob_vector} we have
\begin{equation}\label{lower_bound_eq}
\text{P}_{\pi_l}=\mathbb{E}
 \left[p\left(\left(0,h\right), d_l\left(0,h\right)\right)\right] \ .
\end{equation}
This can be evaluated using \eqref{suc_prob_endtoend_blockn} and \eqref{decision_rule_IRM} for each state $\left(0,h\right)$ with $h \in \mathcal{H}$ and taking the average using the probability mass function $f_\mathcal{H}$.

\subsection{Upper Bound Calculation}
\begin{figure}
	\centering
	\includegraphics[width=8cm]{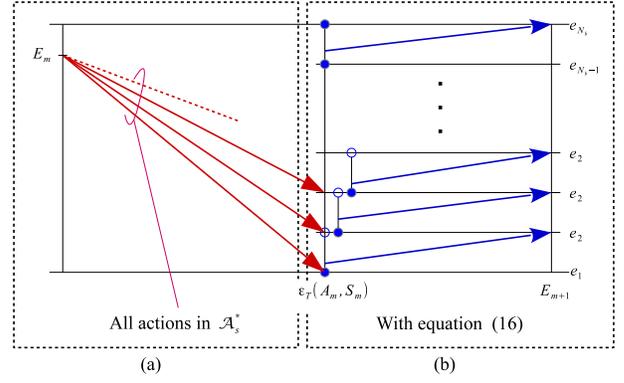}
	\caption{Discretization of the battery energy levels.}
	\label{Modification}
\end{figure}
Although, the state transition of any policy can be modeled by a Markov chain, finding an upper bound using a MDP is involved due to the state space $\mathcal{S}$ is uncountably infinite. Therefore, instead of formulating a MDP for the original system model, we first appropriately modify the system to have a finite state space. We prove that the maximum of the average success probability of the finite state space system gives an upper bound for the maximum of the average success probability of the original system.  To this end, we  discretize the battery energy assuming that there exists a hypothetical energy source in the relay, which injects energy to the battery at the beginning of each block, such that battery energy occupy only predefined $N_b$ number of levels. For the current state $S_m$ and action $A_m$ the residual battery energy for the next block given in \eqref{end_block_energy} is modified by the hypothetical energy source according to
\begin{equation}\label{upper_modi}
E_{m+1} = \begin{cases}
e_{i+1}=\frac{i B}{N_b-1}, \ \text{if} \ \mathcal{E}_{T}\left(A_m,S_m\right) \in \left[\frac{\left(i-1\right)B}{N_b-1} , \frac{ i B}{N_b-1} \right)\\ \quad\quad\quad\quad\quad\quad\quad \text{for each} \ i=1,2,\cdots,N_b-1 \\ e_{N_b}=B,  \quad\quad  \text{otherwise} \ .
\end{cases}
\end{equation}
Each $e_i=\frac{ \left(i-1\right) B}{N_b-1}$ for all $i=1,\cdots,N_b$ denotes the finite battery levels in the battery. According to \eqref{upper_modi}, the hypothetical energy source drives the battery energy to the nearest upper level defined by each $e_i$. This is shown in Fig.~\ref{Modification}b. Thus, the state space has finite number of elements and we denote it by $\mathcal{S}'=\left\{e_1,\cdots,e_{N_b}\right\}\times \mathcal{H}$. We denote a general element of $\mathcal{S}'$ by $s_i$, which are indexed in such a way, that states $\left(e_j, h^{(1)}\right)$ to $\left(e_j, h^{(N_c)}\right)$ map with $s_{(j N_c -N_c +1)}$ to $s_{(j N_c)}$, respectively. Due to the finite nature of the state space, one-step transition probability from the state $S_m$ to state $S_{m+1}$ for any decision rule $d$ can be given in a matrix form according to
\begin{equation}\label{Theta_d}
\Theta_d^{\left(i,j\right)}=\Theta_d\left(s_i,s_j\right) = \Pr\left(S_{m+1}=s_j\ \big| \ S_{m}=s_i \right) \ .
\end{equation}
If the current state is $s_i$ and the residual battery energy determined by the action is $e_j$, the $i$th row of the transition matrix $\Theta_d$ consists of the channel probability values $f_{\mathcal{H}}\left(h^{(1)}\right)$ to $f_{\mathcal{H}}\left(h^{(N_c)}\right)$ from column $N_c\left(j-1\right)+1$ to column $N_c j$. 

Since the state space is finite, for any decision rule $d$, we can define a reward vector $p_d$ in which, each element gives the reward for each state and action defined by the decision rule for the state, i.e. $p_d\left(s_i\right)  = p\big(s_i,d(s_i)\big)$ for all $s_i \in \mathcal{S}' , \ d\in \mathcal{D}$. Using the transition matrix $\Theta_d$ and the reward vector $p_d$ we can write the average success probability of the modified system, in a vector form as \cite{Puterman} 
\begin{equation}\label{lim_inf_avg_suc_prob_vector_modi}
\widetilde{\text{P}}'_\pi =\lim_{M \to \infty} \frac{1}{M} \left[p_{d_1}+\Theta_{d_1} p_{d_2}+\cdots + \prod_{m=1}^{M-1}\Theta_{d_m} p_{d_M}\right] \ .
\end{equation}

The average success probability for the initial state $S_1=s_i$ is given by $\widetilde{\text{P}}'_\pi\left(s_i\right)$, which is the $i$th element of the vector $\widetilde{\text{P}}'_\pi$.  Although the state space $\mathcal{S}'$ is finite, the action space $\mathcal{A}_{s_i}$ for each $s_i\in \mathcal{S}'$ is uncountably infinite for each $s_i$. However, the number of levels of residual battery energy is finite with the modification \eqref{upper_modi}. Thus, we have groups of actions for which the resulting residual battery energy is the same. In fact, it is sufficient to consider a finite action space to find $\underset{\pi \in \Pi}{\text{max}}  \  \widetilde{\text{P}}'_\pi\left(s_i\right)$. This is proved in the next lemma and the proposition.
\begin{lemma}
	\label{Adash}
	For any decision rule $d \in \mathcal{D}$ there exists\break $d' \in \left\{d \ | \ d\left(s\right) \in \mathcal{A}'_s \ \forall s \in \mathcal{S}'  \right\}$ such that $\Theta_d = \Theta_{d'}$, where
	\begin{multline}\label{A_dash}
	\mathcal{A}'_s = \big\{\lambda, u \ | \lambda\in [0,1], \ 0\leqslant u, \ \mathcal{E}_{T}\left(\lambda,u,s\right)=e_i,\\ i=1,2,\cdots,N_b \big\} \ .
	\end{multline}
\end{lemma}
\begin{IEEEproof}
	Channel fading is independent from the decision rule use and we denote $h_{m+1}=h$. Let $E_{m+1}=e_j$ with $j\in \left\{2,\cdots,N_b\right\}$ be the level of residual battery energy resulted from the action $d\left(S_m\right)$ for the state $S_m$. State of the next block is $S_{m+1}=\left(e_j,h\right)$ and we have $\Theta_d\left(S_m,S_{m+1}\right)=f_{\mathcal{H}}\left(h\right)$. In addition, with the aid of (\ref{upper_modi}) it can be seen that the action $d'\left(S_m\right)=\left(\lambda',E'_t\right)$ such that $\mathcal{E}_{T}\left(\lambda',E'_t,S_m\right)=e_{j-1}$ results in the same $E_{m+1}=e_j$. Therefore, we define $\mathcal{A}'_s$ as given in the lemma and thus $d'\left(S_m\right) \in \mathcal{A}'_s$ with $\Theta_{d'}\left(S_m,S_{m+1}\right)=f_{\mathcal{H}}\left(h\right)$, which concludes the proof.
\end{IEEEproof}
Using the following proposition we can further reduce the dimension of $\mathcal{A}_s$ to be finite.
\begin{proposition}
	\label{finite_decision_space}
	For any policy $\pi = \{d_1,d_2,\cdots,\}$ with $d_m\in \mathcal{D}$ for all $m$, there exists a policy $\pi' = \{d_1',d_2',\cdots\}$ with $d_m'\in \tilde{\mathcal{D}}$ for all $m$, such that $\text{P}'_{\pi'}\geqslant \text{P}'_{\pi}$, where
	\begin{equation*}
	\tilde{\mathcal{D}} = \left\{d \ | \ d\left(s\right) \in \mathcal{A}^*_s \ \forall s \in \mathcal{S}' \right\} \subset \mathcal{D} \ ,
	\end{equation*}
	\begin{equation*}
	\mathcal{A}^*_s = \mathcal{A}'_{s,1}\cup \mathcal{A}'_{s,2} \ ,
	\end{equation*}
	\begin{equation*}
	\mathcal{A}'_{s,1} = \left\{\lambda, u \ | \ \left(\lambda, u\right) \in \mathcal{A}'_s, \ \lambda=1 \right\} \ ,
	\end{equation*}
	\begin{equation}\label{A2}
	\mathcal{A}'_{s,2} = \begin{cases}
	\phi \qquad \ \text{if} \ s \in \mathcal{C}_1 \\ \quad\quad\ \ \ \text{otherwise,}\\
	\left\{\lambda, u \ | \ \left(\lambda, u\right) \in \mathcal{A}'_s, \ \lambda = \frac{h P_s - 2 \sigma^2 \gamma_\tau}{h P_s -  \sigma^2 \gamma_\tau} \right\}
	\end{cases}  \, ,
	\end{equation}
	where $\phi$ denotes the empty set.
\end{proposition}
\begin{IEEEproof}
	See Appendix A.
\end{IEEEproof}
The operation of $\mathcal{A}^*_s$ is shown in Fig.~\ref{Modification}a. 

With proposition~\ref{finite_decision_space}, we can claim, that for any policy $\pi \in \Pi$, there exists a policy in $\tilde{\Pi}=\tilde{\mathcal{D}}\times\tilde{\mathcal{D}}\times\tilde{\mathcal{D}}\times\cdots$, which has an average success probability, larger or equal to that of policy $\pi$. Therefore, it is sufficient to restrict our attention to the reduced policy space $\tilde{\Pi}$, when we search for a solution to $\underset{\pi \in \Pi}{\text{max}}  \  \text{P}'_\pi\left(s_i\right)$, which is useful to calculate the upper bound $P_u$ as per the following proposition.
\begin{proposition}\label{supremum_modifed}
	Average success probability in the modified system $\text{P}'_{\pi}$ satisfies, $\underset{\pi}{\text{max}} \ \widetilde{\text{P}}'_{\pi}\left(s_i\right)\geqslant \underset{\pi}{\text{max}}\ \widetilde{\text{P}}_{\pi}\left(s_i\right)$ for all $s_i \in \mathcal{S}'$
\end{proposition}
\begin{IEEEproof}
	See Appendix B.
\end{IEEEproof}
Therefore, the upper bound $P_u$ can be calculated using
\begin{equation}\label{op_prob_reduced}
P_u = \mathbb{E}\left[ \underset{\pi \in \tilde{\Pi}}{\text{max}}  \  \widetilde{\text{P}}'_\pi\big(\left(0,h_1\right)\big)\right] \ .
\end{equation}
Since the state space $\mathcal{S}'$ and the set $\tilde{\mathcal{D}}$ are both finite, the existence of $\underset{\pi \in \tilde{\Pi}}{\text{max}}  \  \text{P}'_\pi\big(s\big)$ for all $s\in \mathcal{S}'$, is guaranteed \cite[Chapter 9]{Puterman}. To evaluate $\underset{\pi \in \tilde{\Pi}}{\text{max}}  \  \text{P}'_\pi\big(s\big)$, we can use a standard average reward policy iteration algorithm, which consists of iterations of following two steps, 
\begin{itemize}
	\item At iteration $n$ ; $\pi_n \leftarrow d_n^\infty$
	\begin{itemize}
		\item \emph{Step-1} ; $\widetilde{\text{P}}'_{\pi_n} \leftarrow Evaluate\_Policy\left(\pi_n\right)$ \ ,
		
		\item \emph{Step-2} ; $d_{n+1} \leftarrow Improve\_Policy\left(\widetilde{\text{P}}'_{\pi_n}\right)$ \ .
	\end{itemize}
\end{itemize} 

The policy iteration algorithm can be initiated with any resource allocation policy $\pi_1 = d_1^\infty$. For the details of the functions $Evaluate\_Policy\left(\pi_n\right)$, $Improve\_Policy\left(\widetilde{\text{P}}'_{\pi_n}\right)$ and the stopping criterion, the reader is referred to \cite[Algorithm 9.2.1]{Puterman}.

\section{Numerical Results}\label{Numerical}
Although our analysis is valid for any finite fading distributions of $\mathcal{H}$ and $\mathcal{G}$, in this section we consider a equiprobable quantization of a unit mean Rayleigh fading \cite{Parastoo} with $N_c=200$ channel states. Simulation results for $\text{P}_{\pi_l}$ in \eqref{lower_bound_eq} are generated by simulating the system with the stationary policy $\pi_l=d_l^\infty$.
\begin{figure}
	\centering
	\includegraphics[width=8cm]{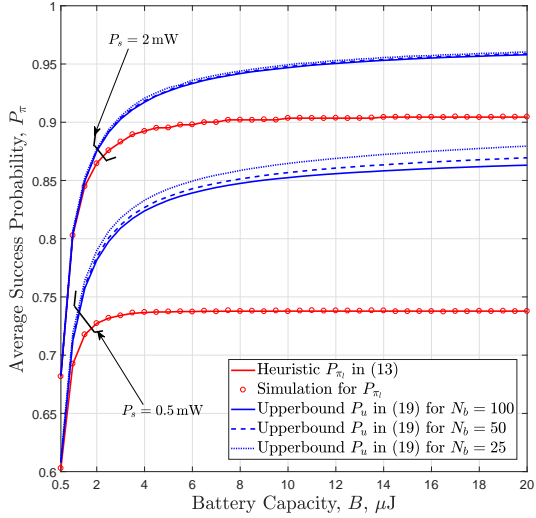}
	\caption{The variation average success probability $\text{P}_\pi$ with the relay battery capacity $B$.}
	\label{Vs_B}
\end{figure}
\begin{figure}
	\centering
	\includegraphics[width=8cm]{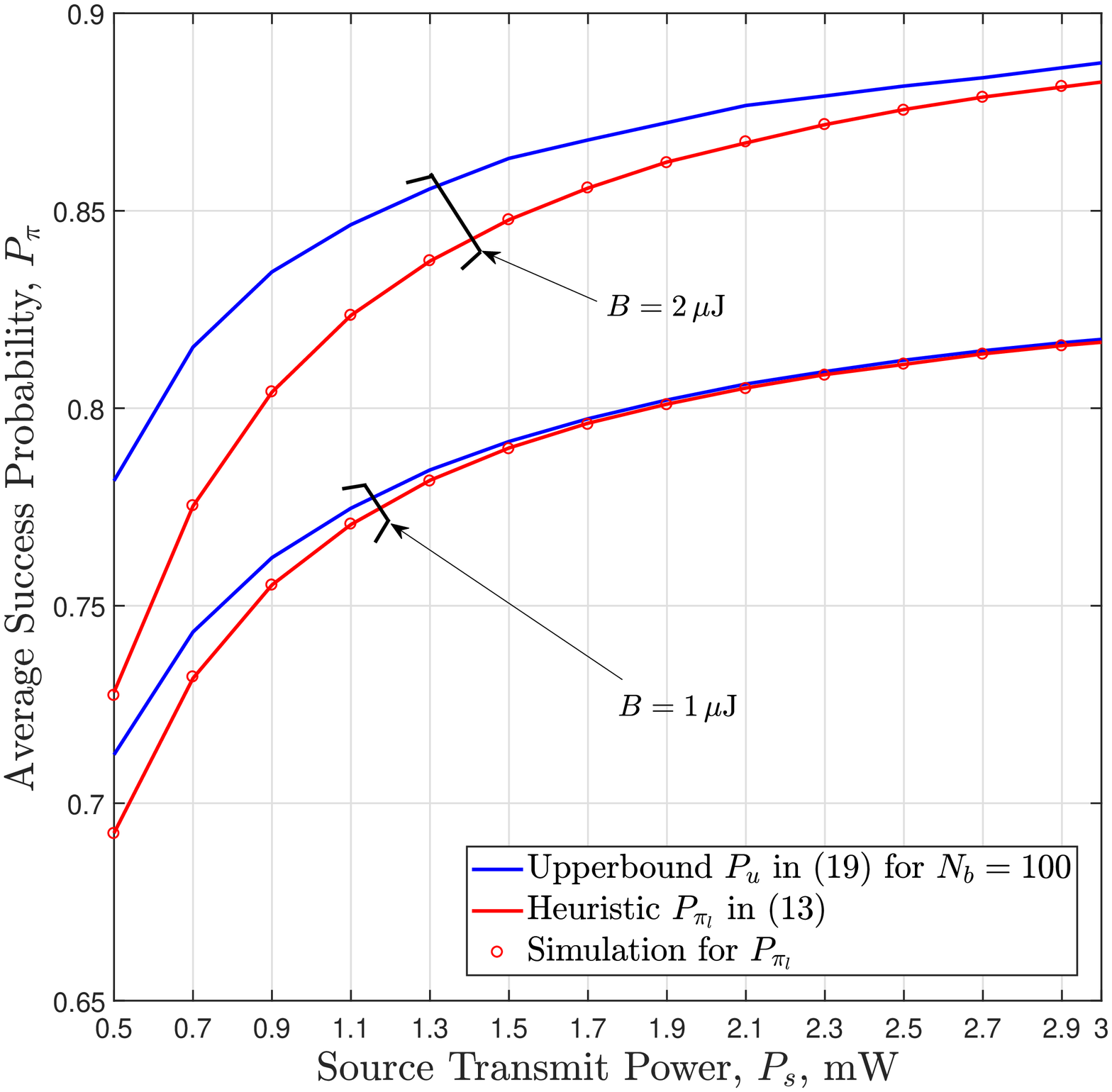}
	\caption{The variation average success probability $\text{P}_\pi$ with the source transmit power $P_s$.}
	\label{Vs_Q}
\end{figure}

Fig.~\ref{Vs_B} shows the variation of $\text{P}_{\pi_l}$ in \eqref{lower_bound_eq} and $P_u$ in \eqref{op_prob_reduced} for difference values of $N_b$ and, with the relay battery capacity $B$, where the source transmit power $P_s=0.5$\,mW and $2$\,mW. Simulation results match with analytical results in \eqref{lower_bound_eq}. As shown in the figure, smaller upper bounds can be obtained with a larger values for $N_b$. The gain of the upper bound from battery capacity $B=B_1$ compared to $B=B_2$ is $100 \times \frac{P_u|_{B=B_1}-P_u|_{B=B_2}}{P_u|_{B=B_2}} \% $. When source transmit power $P_s=2$\,mW, the gain is $29.8\%$ from battery capacity $10$\,$\mu$J compared to $4$\,$\mu$J, whereas the gain is $4.9\%$ from $16$\,$\mu$J compared to $10$\,$\mu$J. For the same increase in the battery capacity, the gain is small. This is also true for $P_s=0.5$\,W. Although a larger battery capacity results in more battery states, occupying a higher battery state is improbable, which explains the diminishing returns in average success probability with battery capacity. The performance gain of $\text{P}_{u}$ compared to $\text{P}_{\pi_l}$ is $100 \times \frac{\text{P}_{\pi^*}-\text{P}_{\pi_l}}{\text{P}_{\pi_l}} \% $. When the source transmit power $P_s=2$\,mW and $B=10$\,$\mu$J the performance gain of $\text{P}_{u}$ is $31\%$ and when the source transmit power $P_s=0.5$\,mW and $B=10$\,$\mu$J the gain is $107.8\%$. 

Fig.~\ref{Vs_Q} shows the variation of $P_u$ and $\text{P}_{\pi_l}$ with the source transmit power $P_s$, for $B=2$\,$\mu$J and $2$\,$\mu$J. Average success probability achieved by the heuristic policy $\pi_l$ gets closer to the upper bound $P_u$ as the source transmit power is increased. This is more noticeable when the battery capacity is small. When the source transmit power $P_s$ is large such that for all $s \in \mathcal{S}$ and $\left(\lambda,u\right) \in \mathcal{A}^*_s$ the half block battery energy is $\mathcal{E}_{\frac{T}{2}}\left(\lambda,s\right)=B$, then for it is optimal to use total battery energy for data transmission to the destination. This makes heuristic policy optimal in this situation, which explains $\text{P}_{\pi_l}$ gets closer to $P_u$ for large $P_s$ or small $B$.

\section{Conclusion}\label{conclusion}
This paper considers SWIPT over a DF relay network with the power-splitting (PS) energy harvesting protocol at the relay. A mathematical framework is presented to investigate the feasibility of the network by evaluating an upper bound of the performance. Numerical results show that performance gain has diminishing returns with battery capacity and the proposed heuristic resource allocation policy achieves a performance close to the upper bound when the source power is large or the relay battery is small. Mathematical framework can be changed to include battery imperfections and power consumption by the information processing circuits and we intend to investigate these in a future work.

\appendix
\subsection{Proof of Proposition~\ref{finite_decision_space}}
We prove that for any policy $\pi$ there exists a policy $\pi'$ as given in the proposition such that $\Theta_{d_m}=\Theta_{d'_m}$ and $p_{d'_m}\geqslant p_{d_m}$ for all $m$, which essentially prove that $\widetilde{\text{P}}'_{\pi'}\geqslant \widetilde{\text{P}}'_{\pi}$ with (\ref{lim_inf_avg_suc_prob_vector_modi}). Using lemma~\ref{Adash}, there exists a decision rule $d'_m$ in $\mathcal{A}'_s$  that gives $\Theta_{d_m}=\Theta_{d'_m}$. The dimension  of $\mathcal{A}'_s$ can be further reduced to have $p_{d'_m}\geqslant p_{d_m}$. We consider cases $\mathcal{C}_1$ and $\mathcal{C}_2$ separately. (i) When $s \in \mathcal{C}_1$, as discussed $p_d\left(s\right)=0$ for all $d$. Therefore, we set $d'\left(s\right)$ to take the corresponding element in $\mathcal{A}'_{s,1}$ such that $\Theta_d\left(s\right)=\Theta_{d'}\left(s\right)$. (ii) When $s \in \mathcal{C}_2$, $\lambda\leqslant \frac{h P_s - 2 \sigma^2 \gamma_\tau}{h P_s -  \sigma^2 \gamma_\tau}$ is feasible for $\mathcal{A}_s$ and for $\mathcal{A}'_s$. We consider two sub cases for $d\left(s\right)=\left(\lambda,u\right)$. (ii.a) When $\lambda\leqslant \frac{h P_s - 2 \sigma^2 \gamma_\tau}{h P_s -  \sigma^2 \gamma_\tau}$. Let the residual battery energy resulted from $d\left(s\right)$ be $e_i$. We set the decision rule $d'\left(s\right)=\left(\lambda', E'_t\right) \in \mathcal{A}'_s$ such that $\lambda'=\frac{h P_s - 2 \sigma^2 \gamma_\tau}{h P_s -  \sigma^2 \gamma_\tau}$ and  $\mathcal{E}_{T}\left(\lambda,u,s\right)=e_i$. It can be shown with (\ref{end_block_energy})  that this makes $E'_t \geqslant u$, which results in $p_{d'}\left(s\right)\geqslant p_{d}\left(s\right)$ in (\ref{suc_prob_endtoend_blockn}). It should be noted that $d'\left(s\right)\in \mathcal{A}'_{s,2}$. (ii.b) When $\lambda > \frac{h P_s - 2 \sigma^2 \gamma_\tau}{h P_s -  \sigma^2 \gamma_\tau}$. In this situation $p_d\left(s\right)=0$. Therefore, we set $d'\left(s\right)$ to take the corresponding element in $\mathcal{A}'_{s,1}$. The new decision rules $d'_m$ take only the elements in $\mathcal{A}'_{s,1}\cup \mathcal{A}'_{s,2}$ and we have $\Theta_{d_m}=\Theta_{d'_m}$ and $p_{d'_m}\geqslant p_{d_m}$ for all $m$, which proves $\widetilde{\text{P}}'_{\pi'}\geqslant \widetilde{\text{P}}'_{\pi}$. This concludes the proof.

\subsection{Proof of Proposition~\ref{supremum_modifed}}
We first compare average success probability over $M$ blocks given in \eqref{lim_avg_suc_prob} for the two systems for a general $M$, where we denote it for the modified system by $\bar{p}'_{\pi,M}$. We use the backward induction method to prove that $\underset{\pi}{\text{max}} \ \bar{p}'_{\pi,M}\geqslant \underset{\pi}{\text{max}}\ \bar{p}_{\pi,M}$, which leads to the results in the proposition. Here, $\bar{p}'_{\pi,M}$ is defined similar to $\bar{p}_{\pi,M}$ in \eqref{lim_avg_suc_prob}

For any given $S_M=\left(E_M,h_{1,M}\right)$ the optimal decision rule that maximize $p\left(S_M,d_M\left(S_M\right)\right)$ denoted by $d^*_M$ uses total energy in the relay battery. Therefore, if the state $S'_M=\left(E'_M,h_{1,M}\right)$ is such that $E'_M>E_M$ then $p\left(S'_M,d^*_M\left(S_M\right)\right)\geqslant p\left(S_M,d^*_M\left(S_M\right)\right)$. For any given $S_{M-1}$ and action $A_{M-1}$ if the original system gives $S_M=\left(E_M,h_{1,M}\right)$, the modified system gives $S'_M=\left(E'_M,h_{1,M}\right)$ with $E'_M\geqslant E_M$. Thus we have
\begin{multline*}
\mathbb{E}_{S_{M-1}}\left[p\left(S_{M-1},A_{M-1}+p\left(S'_M,d^*_M\left(S'_M\right)\right)\right)\right]\geqslant\\ \mathbb{E}_{S_{M-1}}\left[p\left(S_{M-1},A_{M-1}+p\left(S_M,d^*_M\left(S_M\right)\right)\right)\right] \ .
\end{multline*}
Let the two states $S_{M-1}=\left(E_{M-1},h_{1,{M-1}}\right)$ and $S'_{M-1}=\left(E'_{M-1},h_{1,{M-1}}\right)$ be such that $E'_{M-1}>E_{M-1}$ and let the optimal action for $S_{M-1}$ that maximize the sum
\begin{equation*}
\mathbb{E}_{S_{M-1}}\left[p\left(S_{M-1},A_{M-1}+p\left(S'_M,d^*_M\left(S'_M\right)\right)\right)\right]
\end{equation*}
be $A^*_{M-1}$. Since $S'_{M-1}$ has a lager battery energy, with the aid of \eqref{diff_eqn_sol} and \eqref{end_block_energy} it can be seen that the action $A^*_{M-1}$ is feasible for $S'_{M-1}$ and results in a larger $E_{M}$ compared taking the action $A^*_{M-1}$ in $S'_{M-1}$. Thus we have
\begin{multline*}
\mathbb{E}_{S'_{M-1}}\left[p\left(S'_{M-1},A^*_{M-1}+p\left(S'_M,d^*_M\left(S'_M\right)\right)\right)\right]\geqslant\\ \mathbb{E}_{S_{M-1}}\left[p\left(S_{M-1},A^*_{M-1}+p\left(S'_M,d^*_M\left(S'_M\right)\right)\right)\right] \ .
\end{multline*}
Thus the optimal action for $S'_{M-1}$ denoted by $A^{**}_{M-1}$ should satisfy
\begin{multline*}
\mathbb{E}_{S'_{M-1}}\left[p\left(S'_{M-1},A^{**}_{M-1}+p\left(S'_M,d^*_M\left(S'_M\right)\right)\right)\right]\geqslant\\ \mathbb{E}_{S_{M-1}}\left[p\left(S_{M-1},A^*_{M-1}+p\left(S'_M,d^*_M\left(S'_M\right)\right)\right)\right] \ .
\end{multline*} This line of argument can be extended to all the remaining blocks from $M-2$ to $1$, which proves that 
\begin{equation}\label{ap_eq1}
\underset{\pi}{\text{max}} \ \bar{p}'_{\pi,M}\left(s_i\right)\geqslant \underset{\pi}{\text{max}}\ \bar{p}_{\pi,M}\left(s_i\right) \ , \forall \ s_i \in \mathcal{S}' \ .
\end{equation}
To prove $\underset{\pi}{\text{max}} \ \widetilde{\text{P}}'_{\pi}\left(s_i\right)\geqslant \underset{\pi}{\text{max}}\ \widetilde{\text{P}}_{\pi}\left(s_i\right)$, we next prove that $\lim_{M\to \infty}\underset{\pi}{\text{max}} \ \bar{p}'_{\pi,M}\left(s_i\right) = \underset{\pi}{\text{max}} \ \widetilde{\text{P}}'_{\pi}\left(s_i\right)$. With the aid of \eqref{ap_eq1}, we thus have $\underset{\pi}{\text{max}} \ \widetilde{\text{P}}'_{\pi}\left(s_i\right) \geqslant \underset{\pi}{\text{max}} \ \widetilde{\text{P}}_{\pi}\left(s_i\right)$. From the definition of the limit \eqref{lim_inf_avg_suc_prob_vector_modi}, we have that, for a positive real number $\delta>0$ and a policy $\pi$, there exists a natural number $N_\pi$ such that
\begin{equation*}
\widetilde{\text{P}}'_{\pi}\left(s_i\right) - \delta \leqslant \bar{p}'_{\pi,M}\left(s_i\right) \leqslant \widetilde{\text{P}}'_{\pi}\left(s_i\right) +\delta \ ,
\end{equation*}
for all $M>N_\pi$. Let $N=\text{max}\left(N_\pi\right)$, then for all $\pi$ and $M>N$ we have 
\begin{equation*}
\widetilde{\text{P}}'_{\pi}\left(s_i\right) - \delta \leqslant \bar{p}'_{\pi,M}\left(s_i\right) \leqslant \widetilde{\text{P}}'_{\pi}\left(s_i\right) +\delta \ .
\end{equation*}
Therefore, for all $M>N$
\begin{equation*}
\underset{\pi}{\text{max}} \left(\widetilde{\text{P}}'_{\pi}\left(s_i\right) - \delta\right) \leqslant \underset{\pi}{\text{max}}\, \bar{p}'_{\pi,M}\left(s_i\right) \leqslant \underset{\pi}{\text{max}} \left(\widetilde{\text{P}}'_{\pi}\left(s_i\right) + \delta\right) \ .
\end{equation*}
Since $\underset{\pi}{\text{max}} \left(\widetilde{\text{P}}'_{\pi}\left(s_i\right) - \delta\right)=\underset{\pi}{\text{max}} \left(\widetilde{\text{P}}'_{\pi}\left(s_i\right)\right)-\delta$ and $\underset{\pi}{\text{max}} \left(\widetilde{\text{P}}'_{\pi}\left(s_i\right) + \delta\right)=\underset{\pi}{\text{max}} \left(\widetilde{\text{P}}'_{\pi}\left(s_i\right)\right)+\delta$, we have 
\begin{align*}
\lim_{M\to\infty} \underset{\pi}{\text{max}}\, \bar{p}'_{\pi,M}\left(s_i\right) &= \underset{\pi}{\text{max}} \left(\widetilde{\text{P}}'_{\pi}\left(s_i\right)\right)  \\
&= \underset{\pi \in \tilde{\Pi}}{\text{max}} \left(\widetilde{\text{P}}'_{\pi}\left(s_i\right)\right)
\end{align*}
Since $\mathcal{S}'$ and $\mathcal{A}^*_s$ are finite $\underset{\pi \in \tilde{\Pi}}{\text{max}} \left(\widetilde{\text{P}}'_{\pi}\left(s_i\right)\right)$ exists \cite[chapter 9]{Puterman}. This concludes the proof.

\input{output-B.bbl}


\end{document}